\documentclass[10pt]{article}
\usepackage{graphicx}
\usepackage{amsmath}
\usepackage{amssymb}
\usepackage{caption2}
\begin{document}
\bibliographystyle{prsty}
\begin{center}
{\large {\bf \sc{  Analysis of  the vector and axialvector $QQ\bar{Q}\bar{Q}$  tetraquark states with  QCD sum rules }}} \\[2mm]
Zhi-Gang  Wang \footnote{E-mail: zgwang@aliyun.com.  }, Zun-Yan Di   \\
 Department of Physics, North China Electric Power University, Baoding 071003, P. R. China
\end{center}

\begin{abstract}
In this article, we construct  the axialvector-diquark-axialvector-antidiquark type  currents to study both the vector and axialvector  $QQ\bar{Q}\bar{Q}$ tetraquark states with the QCD sum rules, and obtain the masses
$M_{Y(cc\bar{c}\bar{c},1^{+-})} =6.05\pm0.08\,\rm{GeV}$, $M_{Y(cc\bar{c}\bar{c},1^{--})} =6.11\pm0.08\,\rm{GeV}$,
$M_{Y(bb\bar{b}\bar{b},1^{+-})} =18.84\pm0.09\,\rm{GeV}$, $M_{Y(bb\bar{b}\bar{b},1^{--})}  =18.89\pm0.09\,\rm{GeV}$.
The vector tetraquark states lie $40\,\rm{MeV}$ above   the corresponding centroids of the $0^{++}$, $1^{+-}$ and $2^{++}$ tetraquark states, which is a typical  feature  of the  vector tetraquark states consist of four heavy quarks.
\end{abstract}



\section{Introduction}
The exotic  charmonium-like and bottomonium-like  states, such as the     $Z_c(3900)$,   $Z_c(4025)$, $Z_c(4200)$, $Z(4430)$,
$Z_b(10610)$,  $Z_b(10650)$,  are excellent candidates   for  the multiquark states \cite{PDG}.
If they are really tetraquark states, their constituents are two heavy quarks and two light quarks. Up to now, no exotic tetraquark candidate composed of more than two heavy quarks has been reported. Theoretically, there have  been several approaches  to study  the masses
and widths of the exotic states $Y_{Q}$  with quark composition $QQ\bar{Q}\bar{Q}$, such as the non-relativistic potential models \cite{Silvestre-1986,Lloyd-2004,Barnea-2006,Bai-2016,Richard-2017,GuoFK-bbbb},  the  Bethe-Salpeter equation \cite{Heupel-2012}, the constituent diquark model with spin-spin interaction \cite{Berezhnoy-2012,Rosner-2016,Polosa-2018}, the constituent quark model with color-magnetic interaction \cite{Wu-2016}, the (moment) QCD sum rules \cite{Chen-2016,Wang-QQQQ},  etc.  Experimentally, the ATLAS, CMS and LHCb collaborations have measured the cross section for double charmonium
production \cite{Exp-psipsi}, the CMS collaboration  has observed the $\Upsilon$ pair production \cite{Exp-UpsilonUpsilon}. Recently, the LHCb collaboration studied the $\Upsilon\, \mu^+\mu^- $   invariant-mass distribution for a possible exotic tetraquark  state composed of two $b$  quarks and two $\bar{b}$  quarks  based on a data sample of $pp$  collisions recorded with the LHCb detector at center-of-mass energies $\sqrt{s}=7$, $8$ and $13\,\rm{TeV}$ corresponding to an integrated luminosity of $6.3\, \rm{fb}^{-1} $, and observed no  significant excess \cite{LHCb-1806}. The decays  to the final states $\Upsilon\,\mu^+\mu^-$ can take place through  $Y_{b}(0^{++}/2^{++})\to \Upsilon\Upsilon^*/\Upsilon\Upsilon \to \Upsilon\,\mu^+\mu^-$ or $Y_{b}(1^{--})\to \Upsilon\Upsilon^*/\Upsilon\Upsilon \to \Upsilon\,\mu^+\mu^-$.
 In Ref.\cite{Polosa-2018}, Esposito and  Polosa argue that the
partial width for the $Y_{b}(2^{++})\to \Upsilon\,\mu^+\mu^-$ decay is too small to be
currently observed at the LHC. However, if the barrier between the diquark and antidiquark
is very narrow and the tetraquark width is sufficiently
small, the detection of such a state is still possible.

In 2013, the BESIII collaboration studied  the process  $e^+e^- \to \pi^+\pi^-J/\psi$ at a center-of-mass energy of $4.26\,\rm{GeV}$, and observed a structure $Z_c^\pm(3900)$ in the $\pi^\pm J/\psi$ mass spectrum \cite{BES3900}.  Recently, the BESIII collaboration determined   the  spin and parity of the $Z_c^\pm(3900)$ state to  be $J^P = 1^+$ with a statistical
significance larger than $7\sigma$ over other quantum numbers  \cite{JP-BES-Zc3900}. Analogously, there maybe exist a tetraquark state $Y_{c/b}(1^{+-})$ which decays to
the $\eta_c J/\psi$ or $\eta_b\Upsilon$.

The diquarks (or diquark operators) $\varepsilon^{ijk} q^{T}_j C\Gamma q^{\prime}_k$  have  five  structures  in Dirac spinor space, where the $i$, $j$ and $k$ are color indexes, $C\Gamma=C\gamma_5$, $C$, $C\gamma_\mu \gamma_5$,  $C\gamma_\mu $ and $C\sigma_{\mu\nu}$ for the scalar, pseudoscalar, vector, axialvector  and  tensor diquarks, respectively.
The  favorite   diquark configurations are the scalar ($C\gamma_5$) and axialvector ($C\gamma_\mu$) diquark states from the QCD sum rules \cite{WangDiquark,Tang-Diquark,Dosch-Diquark-1989,WangLDiquark}. The QCD sum rules have been extensively applied to study the tetraquark states and molecular states \cite{QCDSR-Tetraquark-Molecule}.
In Refs.\cite{Wang-Huang-Zc3900,Wang-Huang-Zc3900-2}, we study the mass and width of the $Z_c^\pm(3900)$ with the $C\gamma_\mu \otimes\gamma_5 C -C\gamma_5 \otimes\gamma_\mu C $ type current with the QCD sum rules in details, and reproduce the experimental data satisfactorily.
In Ref.\cite{Wang-Y4260-Zc4020}, we study both the vector and axialvector tetraquark states with the $C\gamma_\mu \otimes\gamma_\nu C -C\gamma_\nu \otimes\gamma_\mu C $
type currents, and reproduce the experimental values of the masses of the $Y(4660)$ and $Z_c(4020/4025)$ satisfactorily.
 The double-heavy diquark states $\varepsilon^{ijk} Q^{T}_j C\gamma_5 Q_k$ cannot exist due to the Pauli principle.
 In previous work, we took the double-heavy diquark states $\varepsilon^{ijk} Q^{T}_j C\gamma_\mu Q_k$ as basic constituents to construct the scalar and tensor tetraquark states with the QCD sum rules \cite{Wang-QQQQ}. Now we extend our previous work to study the vector and axialvector tetraquark states $QQ\bar{Q}\bar{Q}$ with the $C\gamma_\mu \otimes\gamma_\nu C -C\gamma_\nu \otimes\gamma_\mu C $ type currents, which are expected to couple potentially to the lowest tetraquark states, especially for the vector tetraquark states.

The article is arranged as follows:  we derive the QCD sum rules for the masses and pole residues of  the vector and axialvector   $QQ\bar{Q}\bar{Q}$ tetraquark states in section 2; in section 3, we present the numerical results and discussions; section 4 is reserved for our conclusion.

\section{QCD sum rules for  the  vector and axialvector tetraquark states}
In the following, we write down  the two-point correlation functions $\Pi_{\mu\nu\alpha\beta}(p)$  in the QCD sum rules,
\begin{eqnarray}
\Pi_{\mu\nu\alpha\beta}(p)&=&i\int d^4x e^{ip \cdot x} \langle0|T\left\{J_{\mu\nu}(x)J_{\alpha\beta}^{\dagger}(0)\right\}|0\rangle \, ,
\end{eqnarray}
where
\begin{eqnarray}
 J_{\mu\nu}(x)&=&\varepsilon^{ijk}\varepsilon^{imn} \Big\{Q^T_j(x)C\gamma_\mu Q_k(x) \bar{Q}_m(x)\gamma_\nu C \bar{Q}^T_n(x)-Q^T_j(x)C\gamma_\nu Q_k(x) \bar{Q}_m(x)\gamma_\mu C \bar{Q}^T_n(x)\Big\}   \, , \nonumber\\
\end{eqnarray}
where the $i$, $j$, $k$, $m$, $n$ are color indexes, the $C$ is the charge conjugation matrix.
As the tetraquark states have many Fock states, we can study the mixing  with the  substitution,
\begin{eqnarray}
 J_{\mu\nu}(x)&\to&\cos\theta J_{\mu\nu}(x)+\sin\theta \widetilde{J}_{\mu\nu}(x)   \, ,
\end{eqnarray}
where the $\theta$ is a mixing angle, the $\widetilde{J}_{\mu\nu}(x)$ is another (or any) tetraquark current with the same quantum numbers as the current $J_{\mu\nu}(x)$.
We can also study the mixing
between the two quark and tetraquark components with the  substitution,
\begin{eqnarray}
J_{\mu\nu}(x)&\to&\cos\theta J_{\mu\nu}(x) +\sin\theta\,\frac{i}{3}\,\langle\bar{Q}Q\rangle\, \bar{Q}(x)\sigma_{\mu\nu}Q(x) \, ,
\end{eqnarray}
where the heavy quark condensate $\langle\bar{Q}Q\rangle=-\frac{1}{12 m_Q}\langle \frac{\alpha_sGG}{\pi}\rangle+\cdots$ \cite{QQ-Conden}. This may be our next work.

At the  phenomenological side, we can insert  a complete set of intermediate hadronic states with
the same quantum numbers as the current operators $J_{\mu\nu}(x)$  into the
correlation functions $\Pi_{\mu\nu\alpha\beta}(p)$ to obtain the hadronic representation
\cite{SVZ79,Reinders85}. After isolating the ground state
contributions of the  axialvector and vector tetraquark states, we get the following results,
\begin{eqnarray}
\Pi_{\mu\nu\alpha\beta}(p)&=&\frac{\lambda_{ Y^+}^2}{M_{Y^+}^2\left(M_{Y^+}^2-p^2\right)}\left(p^2g_{\mu\alpha}g_{\nu\beta} -p^2g_{\mu\beta}g_{\nu\alpha} -g_{\mu\alpha}p_{\nu}p_{\beta}-g_{\nu\beta}p_{\mu}p_{\alpha}+g_{\mu\beta}p_{\nu}p_{\alpha}+g_{\nu\alpha}p_{\mu}p_{\beta}\right) \nonumber\\
&&+\frac{\lambda_{ Y^-}^2}{M_{Y^-}^2\left(M_{Y^-}^2-p^2\right)}\left( -g_{\mu\alpha}p_{\nu}p_{\beta}-g_{\nu\beta}p_{\mu}p_{\alpha}+g_{\mu\beta}p_{\nu}p_{\alpha}+g_{\nu\alpha}p_{\mu}p_{\beta}\right) +\cdots \, \, ,
\end{eqnarray}
where the $Y^+$ and $Y^-$ denote the axialvector  and vector tetraquark states respectively, the  pole residues  $\lambda_{Y^\pm}$  are defined by
\begin{eqnarray}
  \langle 0|J_{\mu\nu}(0)|Y^+(p)\rangle &=& \frac{\lambda_{Y^+}}{M_{Y^+}} \, \varepsilon_{\mu\nu\alpha\beta} \, \varepsilon^{\alpha}p^{\beta}\, , \nonumber\\
 \langle 0|J_{\mu\nu}(0)|Y^-(p)\rangle &=& \frac{\lambda_{Y^-}}{M_{Y^-}} \left(\varepsilon_{\mu}p_{\nu}-\varepsilon_{\nu}p_{\mu} \right)\, ,
\end{eqnarray}
the  $\varepsilon_\mu$ are the polarization vectors of the vector and axialvector tetraquark states.
We can rewrite the correlation functions $\Pi_{\mu\nu\alpha\beta}(p)$ into the following form according to Lorentz covariance,
\begin{eqnarray}
\Pi_{\mu\nu\alpha\beta}(p)&=&\Pi_{Y^+}(p^2)\left(p^2g_{\mu\alpha}g_{\nu\beta} -p^2g_{\mu\beta}g_{\nu\alpha} -g_{\mu\alpha}p_{\nu}p_{\beta}-g_{\nu\beta}p_{\mu}p_{\alpha}+g_{\mu\beta}p_{\nu}p_{\alpha}+g_{\nu\alpha}p_{\mu}p_{\beta}\right) \nonumber\\
&&+\Pi_{Y^-}(p^2)\left( -g_{\mu\alpha}p_{\nu}p_{\beta}-g_{\nu\beta}p_{\mu}p_{\alpha}+g_{\mu\beta}p_{\nu}p_{\alpha}+g_{\nu\alpha}p_{\mu}p_{\beta}\right) \, .
\end{eqnarray}

Now we project out the components $\Pi_{Y^+}(p^2)$ and $\Pi_{Y^-}(p^2)$ by introducing the operators $P_{Y^+}^{\mu\nu\alpha\beta}$ and $P_{Y-}^{\mu\nu\alpha\beta}$,
\begin{eqnarray}
\widetilde{\Pi}_{Y^+}(p^2)&=&p^2\Pi_{Y^+}(p^2)=P_{Y^+}^{\mu\nu\alpha\beta}\Pi_{\mu\nu\alpha\beta}(p) \, , \nonumber\\
\widetilde{\Pi}_{Y^-}(p^2)&=&p^2\Pi_{Y^-}(p^2)=P_{Y^-}^{\mu\nu\alpha\beta}\Pi_{\mu\nu\alpha\beta}(p) \, ,
\end{eqnarray}
where
\begin{eqnarray}
P_{Y^+}^{\mu\nu\alpha\beta}&=&\frac{1}{6}\left( g^{\mu\alpha}-\frac{p^\mu p^\alpha}{p^2}\right)\left( g^{\nu\beta}-\frac{p^\nu p^\beta}{p^2}\right)\, , \nonumber\\
P_{Y^-}^{\mu\nu\alpha\beta}&=&\frac{1}{6}\left( g^{\mu\alpha}-\frac{p^\mu p^\alpha}{p^2}\right)\left( g^{\nu\beta}-\frac{p^\nu p^\beta}{p^2}\right)-\frac{1}{6}g^{\mu\alpha}g^{\nu\beta}\, .
\end{eqnarray}

In the following,  we briefly outline  the operator product expansion for the correlation functions   $\Pi_{\mu\nu\alpha\beta}(p)$ in perturbative QCD.  We  contract the heavy quark fields  with Wick theorem and obtain the results:
\begin{eqnarray}
\Pi_{\mu\nu\alpha\beta}(p)&=&4i\varepsilon^{ijk}\varepsilon^{imn}\varepsilon^{i^{\prime}j^{\prime}k^{\prime}}\varepsilon^{i^{\prime}m^{\prime}n^{\prime}}\int d^4x e^{ip \cdot x}   \nonumber\\
&&\left\{{\rm Tr}\left[ \gamma_{\mu}S^{kk^{\prime}}(x)\gamma_{\alpha} CS^{jj^{\prime}T}(x)C\right] {\rm Tr}\left[ \gamma_{\beta} S^{n^{\prime}n}(-x)\gamma_{\nu} C S^{m^{\prime}mT}(-x)C\right] \right. \nonumber\\
&&-{\rm Tr}\left[ \gamma_{\mu} S^{kk^{\prime}}(x) \gamma_{\beta} CS^{jj^{\prime}T}(x)C\right] {\rm Tr}\left[ \gamma_{\alpha} S^{n^{\prime}n}(-x) \gamma_{\nu}C S^{m^{\prime}mT}(-x)C\right] \nonumber\\
 &&-{\rm Tr}\left[ \gamma_{\nu} S^{kk^{\prime}}(x)\gamma_{\alpha} CS^{jj^{\prime}T}(x)C\right] {\rm Tr}\left[ \gamma_{\beta} S^{n^{\prime}n}(-x)\gamma_{\mu} C S^{m^{\prime}mT}(-x)C\right]  \nonumber\\
 &&\left.+{\rm Tr}\left[ \gamma_{\nu} S^{kk^{\prime}}(x)\gamma_{\beta} CS^{jj^{\prime}T}(x)C\right] {\rm Tr}\left[ \gamma_{\alpha} S^{n^{\prime}n}(-x)\gamma_{\mu} C S^{m^{\prime}mT}(-x)C\right] \right\}\, ,
\end{eqnarray}
where the   $S_{ij}(x)$   is the full $Q$ quark propagator,
 \begin{eqnarray}
S_{ij}(x)&=&\frac{i}{(2\pi)^4}\int d^4k e^{-ik \cdot x} \left\{
\frac{\delta_{ij}}{\!\not\!{k}-m_Q}
-\frac{g_sG^n_{\alpha\beta}t^n_{ij}}{4}\frac{\sigma^{\alpha\beta}(\!\not\!{k}+m_Q)+(\!\not\!{k}+m_Q)
\sigma^{\alpha\beta}}{(k^2-m_Q^2)^2}\right.\nonumber\\
&&\left.+\frac{g_s^2G^n_{\alpha\beta}G^{n\alpha\beta}}{12} \delta_{ij}m_Q \frac{k^2+m_Q\!\not\!{k}}{(k^2-m_Q^2)^4}
+\cdots\right\} \, ,
\end{eqnarray}
and  $t^n=\frac{\lambda^n}{2}$, the $\lambda^n$ is the Gell-Mann matrix \cite{Reinders85}. Then we compute  the integrals both in the coordinate and momentum spaces to obtain the correlation functions $\Pi_{\mu\nu\alpha\beta}(p)$  therefore the QCD spectral densities through dispersion relation,
 \begin{eqnarray}
\rho_A(s)&=&\frac{{\rm Im}\widetilde{\Pi}_{Y^+}(s)}{\pi}\, , \nonumber\\
\rho_V(s)&=&\frac{{\rm Im}\widetilde{\Pi}_{Y^-}(s)}{\pi}\, ,
\end{eqnarray}
where
  \begin{eqnarray}
\widetilde{\Pi}_{Y^+}(p^2)&=&P_{Y^+}^{\mu\nu\alpha\beta}\Pi_{\mu\nu\alpha\beta}(p) \, , \nonumber\\
\widetilde{\Pi}_{Y^-}(p^2)&=&P_{Y^-}^{\mu\nu\alpha\beta}\Pi_{\mu\nu\alpha\beta}(p) \, .
\end{eqnarray}

 We  take the quark-hadron duality below the continuum thresholds  $s_0$ and perform Borel transform  with respect to
the variable $P^2=-p^2$ to obtain  the QCD sum rules:
\begin{eqnarray}\label{QCDSR}
\lambda^2_{Y}\, \exp\left(-\frac{M^2_{Y}}{T^2}\right)= \int_{16m_Q^2}^{s_0} ds \int_{z_i}^{z_f}dz \int_{t_i}^{t_f}dt \int_{r_i}^{r_f}dr\, \rho_{A/V}(s,z,t,r)  \, \exp\left(-\frac{s}{T^2}\right) \, ,
\end{eqnarray}
\begin{eqnarray}
\rho_A(s,z,t,r)&=& \frac{3m_Q^4}{16\pi^6}\left( s-\overline{m}_Q^2\right)^2+\frac{t z m_Q^2}{8\pi^6}\left( s-\overline{m}_Q^2\right)^2\left( 4s-\overline{m}_Q^2\right) \nonumber\\
&&+rtz(1-r-t-z) \frac{s}{16\pi^6}\left( s-\overline{m}_Q^2\right)^2\left( 7s-4\overline{m}_Q^2\right) \nonumber\\
&&+m_Q^2\langle \frac{\alpha_sGG}{\pi}\rangle \left\{-\frac{1}{r^3} \frac{m_Q^4}{12\pi^4}\delta\left( s-\overline{m}_Q^2\right) -\frac{1-r-t-z}{r^2} \frac{m_Q^2}{12\pi^4}\left[1+s\,\delta\left( s-\overline{m}_Q^2\right)\right]\right. \nonumber\\
&&-\frac{tz}{r^3} \frac{m_Q^2}{12\pi^4}\left[1+s\,\delta\left( s-\overline{m}_Q^2\right)\right] -\frac{tz(1-r-t-z)}{r^2} \frac{1}{12\pi^4}\left[4s+s^2\delta\left( s-\overline{m}_Q^2\right)\right]  \nonumber\\
&&\left.+\frac{1}{r^2} \frac{m_Q^2}{4\pi^4}   +\frac{tz}{r^2} \frac{1}{4\pi^4}\left( 2s-\overline{m}_Q^2\right) \right\} \nonumber\\
&&+\langle \frac{\alpha_sGG}{\pi}\rangle \left\{-\frac{m_Q^2}{48\pi^4}\left( 4s-3\overline{m}_Q^2\right)- \frac{r(1-r-t-z)}{16\pi^4}\left( s-\overline{m}_Q^2\right)^2\right. \nonumber\\
&&- \frac{r(1-r-t-z)}{48\pi^4}s\left( 7s-6\overline{m}_Q^2\right)+\frac{1}{rz} \frac{m_Q^4}{48\pi^4} +\frac{t}{r} \frac{m_Q^2}{24\pi^4}\left( 2s-\overline{m}_Q^2\right)\nonumber\\
&&\left.+ \frac{t(1-r-t-z)}{32\pi^4}\left( s-\overline{m}_Q^2\right)^2+ \frac{t(1-r-t-z)}{48\pi^4}s\left( 6s-5\overline{m}_Q^2\right)\right\}\, ,
\end{eqnarray}

\begin{eqnarray}
\rho_V(s,z,t,r)&=& -\frac{3m_Q^4}{16\pi^6}\left( s-\overline{m}_Q^2\right)^2-\frac{t z m_Q^2}{8\pi^6}\left( s-\overline{m}_Q^2\right)^3 \nonumber\\
&&+rtz(1-r-t-z) \frac{s}{16\pi^6}\left( s-\overline{m}_Q^2\right)^2\left( 7s-4\overline{m}_Q^2\right) \nonumber\\
&&+m_Q^2\langle \frac{\alpha_sGG}{\pi}\rangle \left\{\frac{1}{r^3} \frac{m_Q^4}{12\pi^4}\delta\left( s-\overline{m}_Q^2\right)+\frac{1-r-t-z}{r^2} \frac{m_Q^2}{12\pi^4}\right. \nonumber\\
&&+\frac{tz}{r^3} \frac{m_Q^2}{12\pi^4} -\frac{tz(1-r-t-z)}{r^2} \frac{1}{12\pi^4}\left[4s+s^2\delta\left( s-\overline{m}_Q^2\right)\right]  \nonumber\\
&&\left.-\frac{1}{r^2} \frac{m_Q^2}{4\pi^4}   -\frac{tz}{r^2} \frac{1}{4\pi^4}\left( s-\overline{m}_Q^2\right) \right\} \nonumber\\
&&+\langle \frac{\alpha_sGG}{\pi}\rangle \left\{\frac{m_Q^2}{48\pi^4}\left( 5s-3\overline{m}_Q^2\right)+ \frac{r(1-r-t-z)}{16\pi^4}\left( s-\overline{m}_Q^2\right)^2\right. \nonumber\\
&&+ \frac{r(1-r-t-z)}{48\pi^4}s\left( 7s-6\overline{m}_Q^2\right)-\frac{1}{rz} \frac{m_Q^4}{48\pi^4} -\frac{t}{r} \frac{m_Q^2}{24\pi^4}\left( s-\overline{m}_Q^2\right)\nonumber\\
&&\left.- \frac{t(1-r-t-z)}{32\pi^4}\left( s-\overline{m}_Q^2\right)^2- \frac{t(1-r-t-z)}{48\pi^4}s\left( s-\overline{m}_Q^2\right)\right\}\, ,
\end{eqnarray}
where
\begin{eqnarray}
\overline{m}_Q^2&=&\frac{m_Q^2}{r}+\frac{m_Q^2}{t}+\frac{m_Q^2}{z}+\frac{m_Q^2}{1-r-t-z}\, , \nonumber \\
r_{f/i}&=&\frac{1}{2}\left\{1-z-t \pm \sqrt{(1-z-t)^2-4\frac{1-z-t}{\hat{s}-\frac{1}{z}-\frac{1}{t}}}\right\} \, ,\nonumber\\
t_{f/i}&=&\frac{1}{2\left( \hat{s}-\frac{1}{z}\right)}\left\{ (1-z)\left( \hat{s}-\frac{1}{z}\right)-3 \pm \sqrt{ \left[  (1-z)\left( \hat{s}-\frac{1}{z}\right)-3\right]^2-4 (1-z)\left( \hat{s}-\frac{1}{z}\right)  }\right\}\, ,\nonumber\\
z_{f/i}&=&\frac{1}{2\hat{s}}\left\{ \hat{s}-8 \pm \sqrt{\left(\hat{s}-8\right)^2-4\hat{s}  }\right\}\, ,
\end{eqnarray}
and $\hat{s}=\frac{s}{m_Q^2}$.

 We derive   Eq.\eqref{QCDSR} with respect to  $\tau=\frac{1}{T^2}$, then eliminate the
 pole residues $\lambda_{Y}$, and  obtain the QCD sum rules for
 the masses of the vector   and axialvector  $QQ\bar{Q}\bar{Q}$ tetraquark states,
 \begin{eqnarray}
 M^2_{Y}&=&- \frac{\frac{d}{d \tau}\int_{16m_Q^2}^{s_0} ds \int_{z_i}^{z_f}dz \int_{t_i}^{t_f}dt \int_{r_i}^{r_f}dr\, \rho(s,z,t,r)  \, \exp\left(-\tau s\right)}{\int_{16m_Q^2}^{s_0} ds \int_{z_i}^{z_f}dz \int_{t_i}^{t_f}dt \int_{r_i}^{r_f}dr\, \rho(s,z,t,r)  \, \exp\left(-\tau s\right)}\, .
\end{eqnarray}

\section{Numerical results and discussions}

We take the gluon condensate  to be the standard value
\cite{SVZ79,Reinders85,ColangeloReview}, and  take the $\overline{MS}$ masses $m_{c}(m_c)=(1.28\pm0.03)\,\rm{GeV}$ and $m_{b}(m_b)=(4.18\pm0.03)\,\rm{GeV}$
 from the Particle Data Group \cite{PDG}.
We take into account
the energy-scale dependence of  the  $\overline{MS}$ masses from the renormalization group equation,
 \begin{eqnarray}
m_c(\mu)&=&m_c(m_c)\left[\frac{\alpha_{s}(\mu)}{\alpha_{s}(m_c)}\right]^{\frac{12}{25}} \, ,\nonumber\\
m_b(\mu)&=&m_b(m_b)\left[\frac{\alpha_{s}(\mu)}{\alpha_{s}(m_b)}\right]^{\frac{12}{23}} \, ,\nonumber\\
\alpha_s(\mu)&=&\frac{1}{b_0t}\left[1-\frac{b_1}{b_0^2}\frac{\log t}{t} +\frac{b_1^2(\log^2{t}-\log{t}-1)+b_0b_2}{b_0^4t^2}\right]\, ,
\end{eqnarray}
  where $t=\log \frac{\mu^2}{\Lambda^2}$, $b_0=\frac{33-2n_f}{12\pi}$, $b_1=\frac{153-19n_f}{24\pi^2}$,
   $b_2=\frac{2857-\frac{5033}{9}n_f+\frac{325}{27}n_f^2}{128\pi^3}$,    $\Lambda=210\,\rm{MeV}$, $292\,\rm{MeV}$  and  $332\,\rm{MeV}$ for the flavors
   $n_f=5$, $4$ and $3$, respectively   \cite{PDG}.

In Ref.\cite{Wang-QQQQ}, we study  the energy scale dependence of the predicated masses of the scalar and tensor $QQ\bar{Q}\bar{Q}$ tetraquark states with the QCD sum rules in details. The predicted tetraquark   masses decrease  monotonously and slowly with increase of the energy scales, the QCD sum rules  are stable with variations of the Borel parameters at the energy scales $1.2\,{\rm{GeV}}<\mu<2.2\,{\rm GeV}$ and $2.5\,{\rm{GeV}}<\mu<3.3\,{\rm GeV}$ for the $cc\bar{c}\bar{c}$ and $bb\bar{b}\bar{b}$ tetraquark states, respectively.  At the energy scales  $\mu=2.0\,\rm{GeV}$ and $3.1\,\rm{GeV}$, the relation $\sqrt{s_0}=M_{ gr}+0.5\,\rm{GeV}$ is satisfied, where the $gr$ denotes the ground state tetraquark states, the optimal energy scales of the QCD spectral densities are $\mu=2.0\,\rm{GeV}$ and $3.1\,\rm{GeV}$ for the $cc\bar{c}\bar{c}$ and $bb\bar{b}\bar{b}$ tetraquark states, respectively
 \cite{Wang-QQQQ}. In this article, we choose the same energy scales for the $cc\bar{c}\bar{c}$ and $bb\bar{b}\bar{b}$ tetraquark states, respectively, which work well.

In the QCD sum rules, we usually take the continuum threshold parameters as $\sqrt{s_0}=M_{gr}+(0.4\sim 0.6)\,\rm{GeV}$ for the conventional mesons, where  the $gr$ denotes the ground states.
  Experimentally, the energy gaps $M_{\psi^\prime}-M_{J/\psi}=589\,\rm{MeV}$,   $M_{\eta_c^\prime}-M_{\eta_c}=656\,\rm{MeV}$,
  $M_{\Upsilon^\prime}-M_{\Upsilon}=563\,\rm{MeV}$ and    $M_{\eta_b^\prime}-M_{\eta_b}=600\,\rm{MeV}$ from the Particle Data Group \cite{PDG}.
   The QCD sum rules support assigning  the $Z_c(3900)$ and $Z(4430)$  to be  the ground state and the first radial excited state of the axial-vector  tetraquark states with $J^{PC}=1^{+-}$, respectively,  and assigning the  $X(3915)$ and $X(4500)$ to be the ground state and the first radial excited state of the scalar $cs\bar{c}\bar{s}$ tetraquark states with $J^{PC}=0^{++}$, respectively \cite{Wang4430,Wang4500}.
  The mass gaps are $M_{Z(4430)}-M_{Z_c(3900)}=576\,\rm{MeV}$ and $M_{X(4500)}-M_{X(3915)}=588\,\rm{MeV}$, which also satisfy the relation $\sqrt{s_0}=M_{gr}+(0.4\sim 0.6)\,\rm{GeV}$.
 In this article, we take the relation $\sqrt{s_0}=M_{gr}+(0.4\sim 0.6)\,\rm{GeV}$ as a constraint, and search for the optimal continuum thresholds  $s_0$.

 We search for the optimal  Borel parameters $T^2$ and continuum threshold
parameters $s_0$  to satisfy the  two  criteria of the QCD sum rules:  pole dominance at the phenomenological side and
convergence of the operator product expansion at the QCD side.
The resulting Borel parameters, continuum threshold parameters, energy scales, pole contributions are shown explicitly in Table 1. From the Table, we can see that the
pole contributions are about $(45-60)\%$, the same as that for the scalar and tensor tetraquark states \cite{Wang-QQQQ}, the pole dominance at the phenomenological side is well satisfied.

In the Borel windows, the dominant contributions come from the perturbative terms, the  contributions of the gluon condensate are about $-15\%$, $-3\%$, $-8\%$ and $-2\%$ for the tetraquark states $cc\bar{c}\bar{c}(1^{+-})$, $cc\bar{c}\bar{c}(1^{--})$, $bb\bar{b}\bar{b}(1^{+-})$ and
$bb\bar{b}\bar{b}(1^{--})$, respectively,  the operator product expansion is well convergent. As the dominant contributions come from the perturbative terms, perturbative $\mathcal{O}(\alpha_s)$ corrections amount to multiplying the perturbative terms by a factor $\kappa$, which can be absorbed into the pole residues and cannot impair the predicted masses remarkably.  In the QCD sum rules for the tetraquark states, we usually carry out the
operator product expansion to the vacuum condensates  up to dimension-10  and
assume vacuum saturation for the  higher dimension vacuum condensates \cite{Wang-Huang-Zc3900,Wang-Y4260-Zc4020}. As the vacuum condensates are vacuum  expectations of the quark and gluon operators, we take the truncation  $i\leq 1$ in a consistent way,
the operators of the orders $\mathcal{O}( \alpha_s^{i})$ with $i> 1$ are  discarded, i.e. we take into account the terms
 $\langle\bar{q}q\rangle$, $\langle\frac{\alpha_{s}GG}{\pi}\rangle$, $ \langle\bar{q}g_s\sigma Gq\rangle$,
$ \langle\bar{q}q\rangle^2$, $\langle\bar{q}q\rangle\langle\frac{\alpha_{s}GG}{\pi}\rangle$, $\langle\bar{q}q\rangle\langle\bar{q}g_s\sigma Gq\rangle$,
$\langle\bar{q}g_s\sigma Gq\rangle^2$, $\langle\bar{q}q\rangle^2\langle\frac{\alpha_{s}GG}{\pi}\rangle$
 \cite{Wang-Huang-Zc3900,Wang-Y4260-Zc4020}.  In this article, only gluon condensates have contributions.
Now the two    criteria of the QCD sum rules  are all satisfied, we expect to make reasonable predictions.

\begin{table}
\begin{center}
\begin{tabular}{|c|c|c|c|c|c|c|c|}\hline\hline
                            &$T^2(\rm{GeV}^2)$  &$s_0(\rm{GeV}^2)$ &$\mu(\rm{GeV})$ &pole         &$M_{Y}(\rm{GeV})$    &$\lambda_{Y}(\rm{GeV}^5)$ \\ \hline
$cc\bar{c}\bar{c}(1^{+-})$  &$4.5-4.9$          &$43\pm1$          &$2.0$           &$(46-61)\%$  &$6.05\pm0.08$        &$(2.97\pm0.44)\times10^{-1}$  \\
$cc\bar{c}\bar{c}(1^{--})$  &$4.2-4.6$          &$44\pm1$          &$2.0$           &$(46-62)\%$  &$6.11\pm0.08$        &$(1.82\pm0.33)\times10^{-1}$  \\
$bb\bar{b}\bar{b}(1^{+-})$  &$13.3-13.9$        &$374\pm3$         &$3.1$           &$(48-60)\%$  &$18.84\pm0.09$       &$5.45\pm1.01$  \\
$bb\bar{b}\bar{b}(1^{--})$  &$11.7-12.3$        &$376\pm3$         &$3.1$           &$(47-60)\%$  &$18.89\pm0.09$       &$1.64\pm0.36$  \\  \hline
\end{tabular}
\end{center}
\caption{ The Borel parameters, continuum threshold parameters, energy scales,  pole contributions, masses and pole residues of the $QQ\bar{Q}\bar{Q}$ tetraquark states. }
\end{table}

We take into account all uncertainties of the input parameters, and obtain the values of the ground state masses and pole residues, which are also shown explicitly in Table 1. From Table 1, we can see that  the constraint $\sqrt{s_0}=M_{gr}+(0.4\sim 0.6)\,\rm{GeV}$ is also satisfied. In Figs.1-2, we plot the masses and pole residues with variations of the Borel parameters at larger intervals   than the  Borel windows shown in Table 1.
From Figs.1-2, we can see that the predicted masses and pole residues are rather stable with variations of the Borel parameters. The uncertainties originate from the Borel parameters in the Borel windows are very small, there appear Borel platforms in the Borel windows.

In Fig.3, we plot the predicted masses $M_{Y}$  with variations of the energy scales $\mu$ for the central values of the input parameters shown in Table 1.
 From the figure, we can see that the masses $M_{Y}$ decrease monotonously and slowly with increase of the energy scales  $\mu$. In this article, we choose the same energy scales as the corresponding ones for the $0^{++}$ and $2^{++}$ tetraquark states $QQ\bar{Q}\bar{Q}$, the uncertainties originate from the energy scales cannot  impair the predicative ability remarkably.

In Table 2, we present all the masses of the $0^{++}$, $1^{+-}$, $2^{++}$ and $1^{--}$ tetraquark states $QQ\bar{Q}\bar{Q}$ from the QCD sum rules in Ref.\cite{Wang-QQQQ} and in the present  work.  If we take the central values, the mass-splittings among  the spin-multiplets are not consistent with the simple spin-spin interaction $C\,\vec{S}_1 \cdot \vec{S}_2$ between diquarks, where the $C$ is a fitted constant, the $\vec{S}_{1}$ and $\vec{S}_{2}$ are the spins of the diquark and antidiquark, respectively. The spin-spin interactions among the quarks can be written as $\frac{C}{m_i m_j}\vec{s}_1\cdot \vec{s}_2$, where the $C$ is a fitted constant, the $\vec{s}_{1}$ and $\vec{s}_{2}$ are the spins of the quark and antiquark, respectively. Although the mass-splittings (if the central values are taken) among  the spin-multiplets are also not consistent with the interaction $\frac{C}{m_i m_j}\vec{s}_1\cdot \vec{s}_2$ quantitatively, they are reasonable qualitatively, the $cc\bar{c}\bar{c}$ tetraquark states have larger mass-splittings  according  to the factor $\frac{1}{m_i m_j}$. Considering the uncertainties of the predicted tetraquark masses, we cannot draw the conclusion that the mass-splittings are not consistent with the spin-spin interactions indeed.

The vector tetraquark states lie  $40\,\rm{MeV}$ above the corresponding centroids of the $0^{++}$, $1^{+-}$ and $2^{++}$ tetraquark states. Naively, we expect that  an additional P-wave costs about $500\,\rm{MeV}$, which is much larger than the energy gap $40\,\rm{MeV}$. This maybe a typical  feature  of the  vector tetraquark states consist of four heavy quarks.
 The calculations based on the QCD sum rules indicate that  the $C\gamma_\mu \otimes\gamma_\nu C -C\gamma_\nu \otimes\gamma_\mu C $ type vector tetraquark state $cq\bar{c}\bar{q}$ has a mass $4.66\pm0.09 \, \rm{GeV}$, the $C \otimes \gamma_\mu C$ type vector tetraquark state $cs\bar{c}\bar{s}$ has a mass
 $4.66\pm0.09 \, \rm{GeV}$, which are all consistent with the $Y(4660/4630)$, the   $C\gamma_5 \otimes \gamma_5\gamma_\mu C$  type vector tetraquark state $cq\bar{c}\bar{q}$ has a mass  $4.34\pm0.08\, \rm{GeV}$, which is consistent with the  $Y(4360/4320)$   \cite{Wang-Y4260-Zc4020,Wang-Vector-2018}. The energy gap between the  vector and
axialvector hidden-charm tetraquark states is about or larger than $440\,\rm{MeV}$, which is much larger than $40\,\rm{MeV}$.

In Ref.\cite{Wang-Y4260-Zc4020}, we choose the current,
\begin{eqnarray}
\eta_{\mu\nu}(x)&=&\frac{\varepsilon^{ijk}\varepsilon^{imn}}{\sqrt{2}}\Big\{u^T_j(x) C\gamma_\mu c_k(x) \bar{d}_m(x) \gamma_\nu C \bar{c}^T_n(x)  -u^T_j(x) C\gamma_\nu c_k(x) \bar{d}_m(x) \gamma_\mu C \bar{c}^T_n(x) \Big\} \, , \nonumber\\
 \end{eqnarray}
which has the same structure as the current $J_{\mu\nu}(x)$ in the present work. In the QCD sum rules for the vector tetraquark state $Y(4660)$, the pole contribution is $(46-64)\%$ \cite{Wang-Y4260-Zc4020}, while in the  QCD sum rules for the vector $QQ\bar{Q}\bar{Q}$ tetraquark states, the pole contributions are $(46-62)\%$ and
$(47-60)\%$, we can draw the conclusion tentatively that  the same pole contributions lead to quite  different mass-splittings  between the vector and axialvector tetraquark states.    The four-heavy tetraquark states maybe have typical features due to  absence of light quark contributions.

\begin{table}
\begin{center}
\begin{tabular}{|c|c|c|c|c|c|c|c|}\hline\hline
                             &$M_{Y}(\rm{GeV})$  &Centroids (\rm{GeV})      \\ \hline
$cc\bar{c}\bar{c}(0^{++})$   &$5.99\pm0.08$      &$6.07\pm0.08$            \\
$cc\bar{c}\bar{c}(1^{+-})$   &$6.05\pm0.08$      &                         \\
$cc\bar{c}\bar{c}(2^{++})$   &$6.09\pm0.08$      &                        \\ \hline
$bb\bar{b}\bar{b}(0^{++})$   &$18.84\pm0.09$     &$18.85\pm0.09$           \\
$bb\bar{b}\bar{b}(1^{+-})$   &$18.84\pm0.09$     &                         \\
$bb\bar{b}\bar{b}(2^{++})$   &$18.85\pm0.09$     &                         \\ \hline
$cc\bar{c}\bar{c}(1^{--})$   &$6.11\pm0.08$      &                         \\
$bb\bar{b}\bar{b}(1^{--})$   &$18.89\pm0.09$     &                         \\  \hline \hline
 \end{tabular}
\end{center}
\caption{ The  masses of the tetraquark states $QQ\bar{Q}\bar{Q}$ from the QCD sum rules. }
\end{table}

 The values of the thresholds are $2M_{\eta_c}=5966.8\,\rm{MeV}$, $2M_{J/\psi}=6193.8\,\rm{MeV}$, $M_{\eta_c}+M_{J/\psi}=6080.3\,\rm{MeV}$, $2M_{\eta_b}=18798.0\,\rm{MeV}$, $2M_{\Upsilon}=18920.6\,\rm{MeV}$, $M_{\eta_b}+M_{\Upsilon}=18859.3\,\rm{MeV}$ from the Particle Data Group \cite{PDG}.
 The decays
\begin{eqnarray}
Y(cc\bar{c}\bar{c},1^{+-}) &\to& \eta_c J/\psi \to \mu^+\mu^- +{\rm light\,hadrons}\, ,\nonumber\\
Y(bb\bar{b}\bar{b},1^{+-}) &\to& \eta_b \Upsilon \to \mu^+\mu^- +{\rm light\,hadrons}\, ,\nonumber\\
Y(bb\bar{b}\bar{b},1^{--}) &\to& \Upsilon \Upsilon \to \mu^+\mu^- \mu^+\mu^-\, ,
\end{eqnarray}
can take place with very small phase spaces.
The decays
 \begin{eqnarray}
X(cc\bar{c}\bar{c},1^{--}) &\to& J/\psi {J/\psi}^* \to \mu^+\mu^- \mu^+\mu^- \, ,
\end{eqnarray}
can take place through the virtual $J/\psi^*$.
We can search for the $Y(cc\bar{c}\bar{c},1^{+-}/1^{--})$ and $Y(bb\bar{b}\bar{b},1^{+-}/1^{--}) $ in the mass spectrum of the $\mu^+\mu^- \mu^+\mu^-$ or $\mu^+\mu^- +{\rm light\,hadrons}$ in the future.

\begin{figure}
 \centering
 \includegraphics[totalheight=5cm,width=7cm]{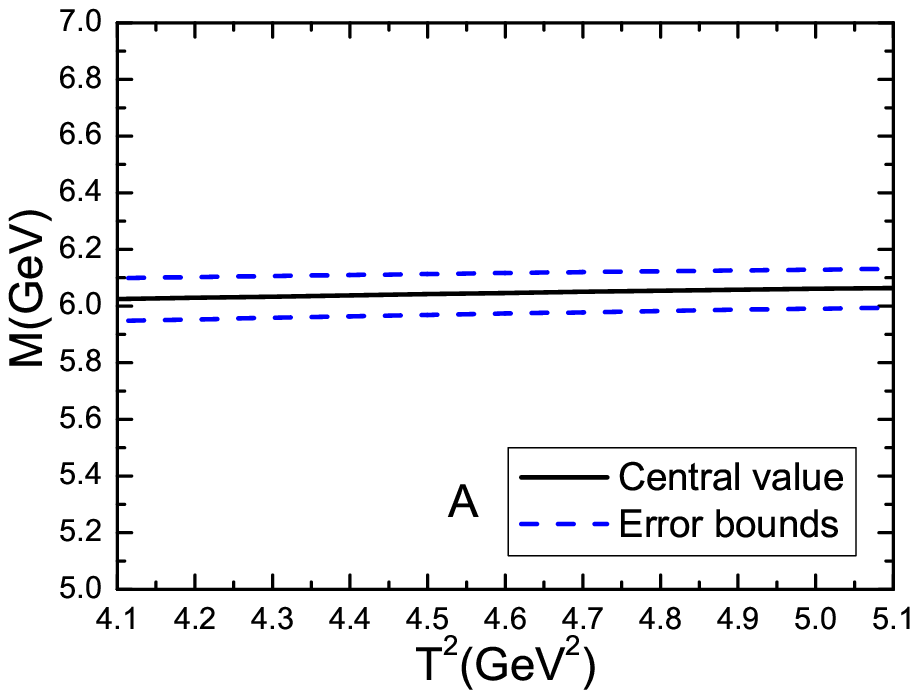}
 \includegraphics[totalheight=5cm,width=7cm]{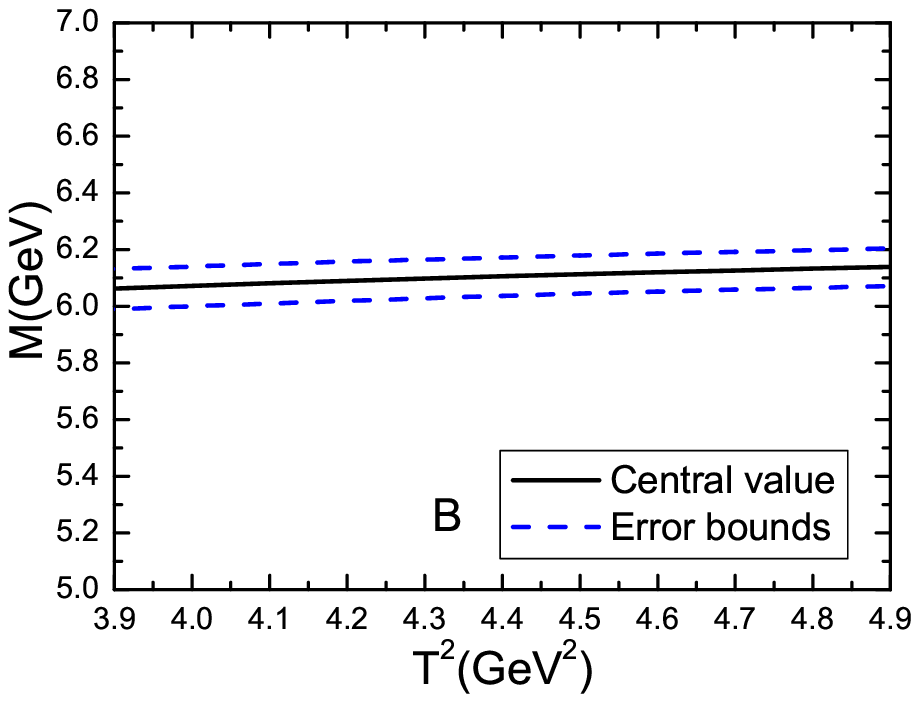}
 \includegraphics[totalheight=5cm,width=7cm]{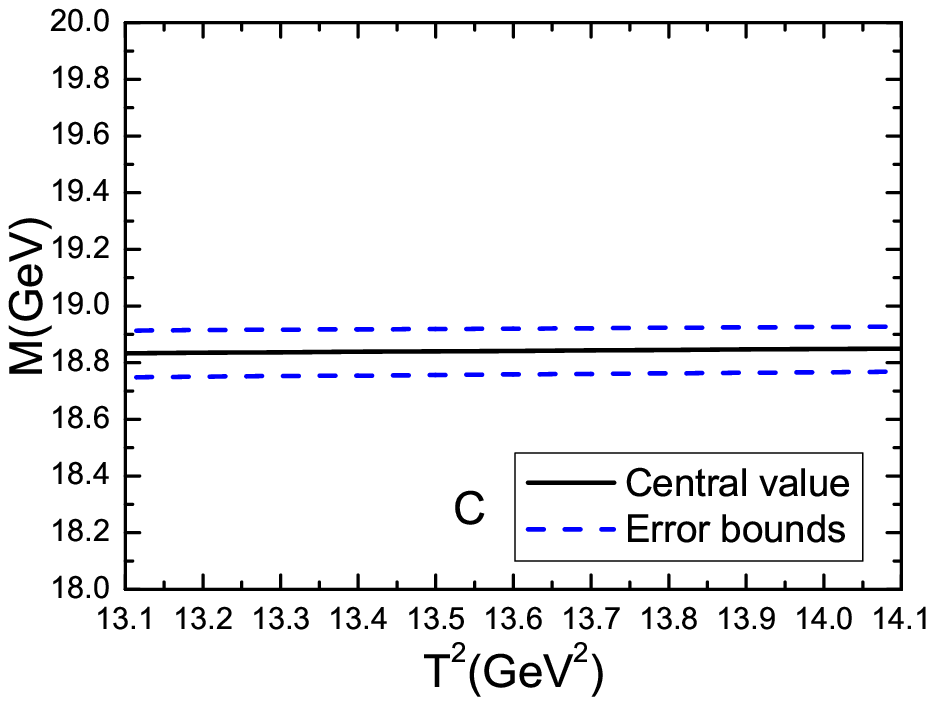}
 \includegraphics[totalheight=5cm,width=7cm]{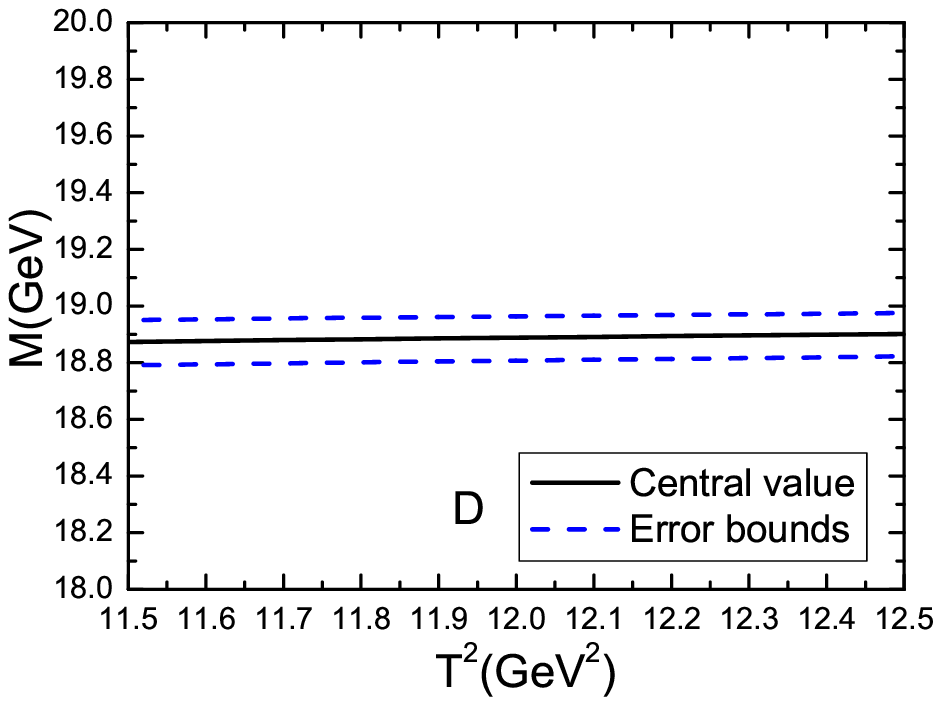}
         \caption{ The masses   of the tetraquark states  with variations of the Borel parameters $T^2$,  where the $A$, $B$, $C$ and $D$ denote the $cc\bar{c}\bar{c}(1^{+-})$, $cc\bar{c}\bar{c}(1^{--})$, $bb\bar{b}\bar{b}(1^{+-})$ and $bb\bar{b}\bar{b}(1^{--})$, respectively.  }
\end{figure}

\begin{figure}
 \centering
 \includegraphics[totalheight=5cm,width=7cm]{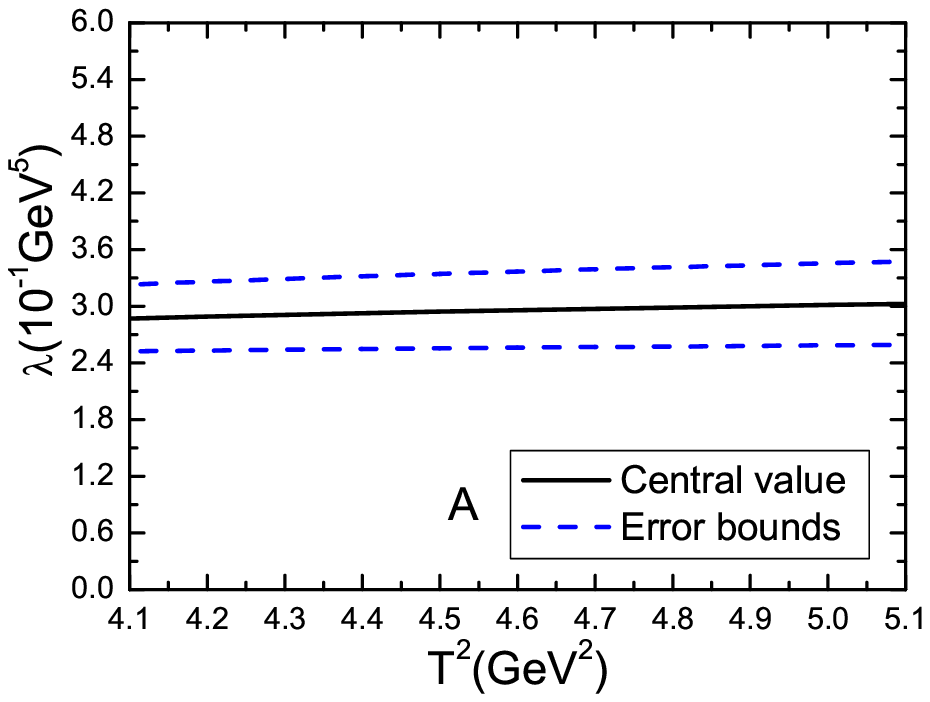}
 \includegraphics[totalheight=5cm,width=7cm]{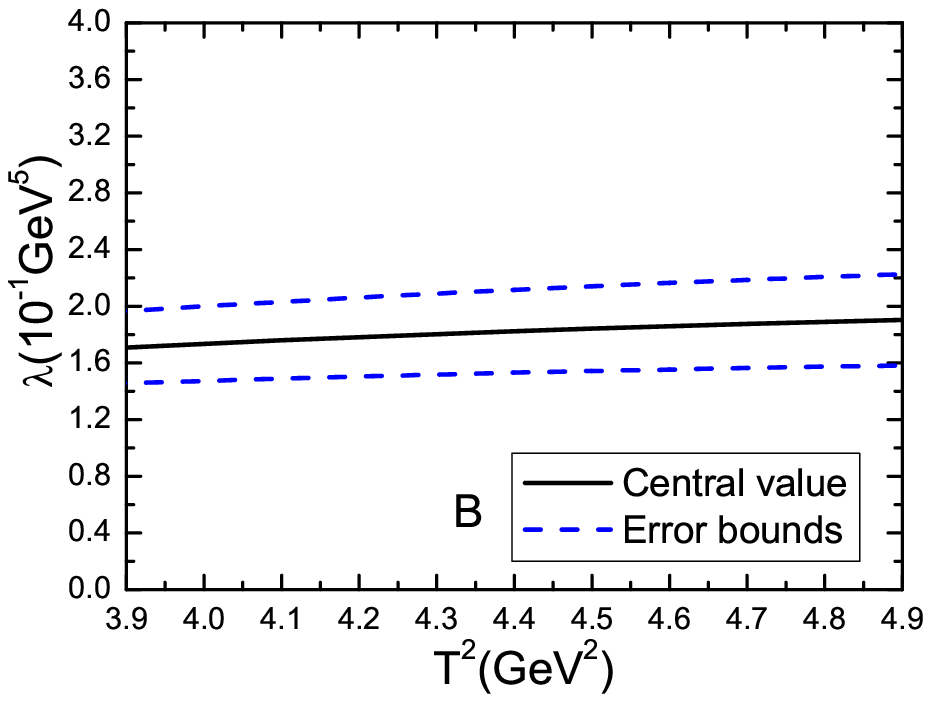}
 \includegraphics[totalheight=5cm,width=7cm]{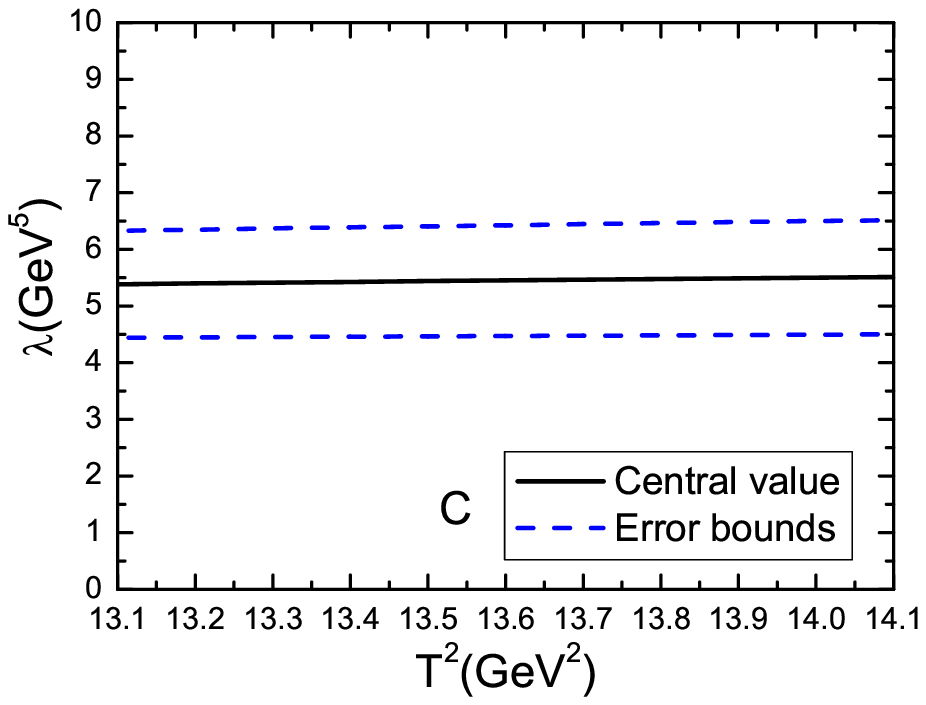}
 \includegraphics[totalheight=5cm,width=7cm]{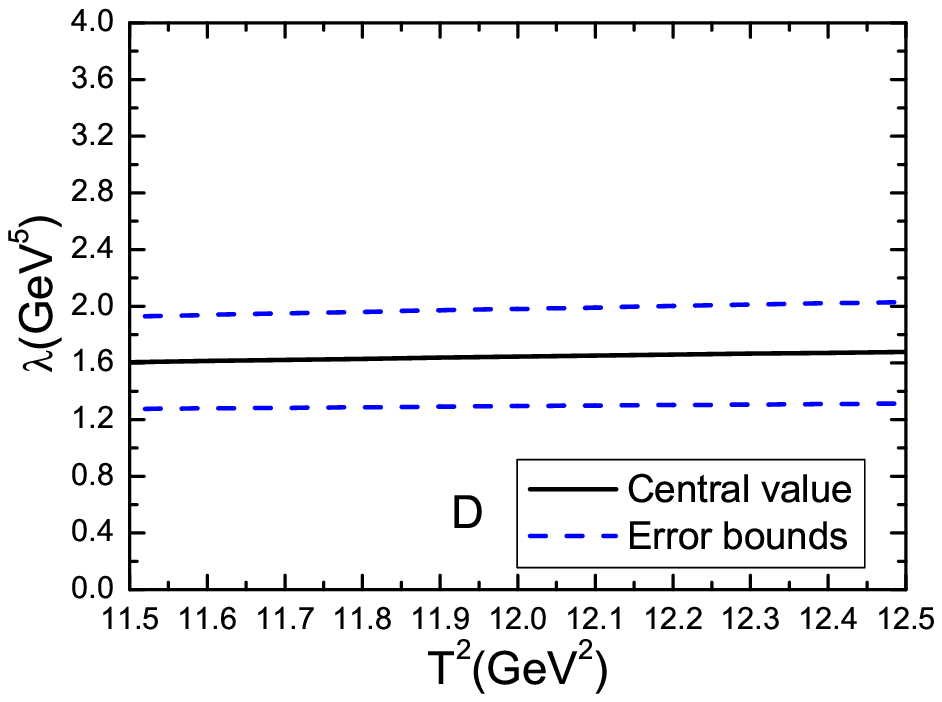}
         \caption{ The pole residues    of the tetraquark states  with variations of the Borel parameters $T^2$,  where the $A$, $B$, $C$ and $D$ denote the $cc\bar{c}\bar{c}(1^{+-})$, $cc\bar{c}\bar{c}(1^{--})$, $bb\bar{b}\bar{b}(1^{+-})$ and $bb\bar{b}\bar{b}(1^{--})$, respectively.  }
\end{figure}

\begin{figure}
 \centering
 \includegraphics[totalheight=5cm,width=7cm]{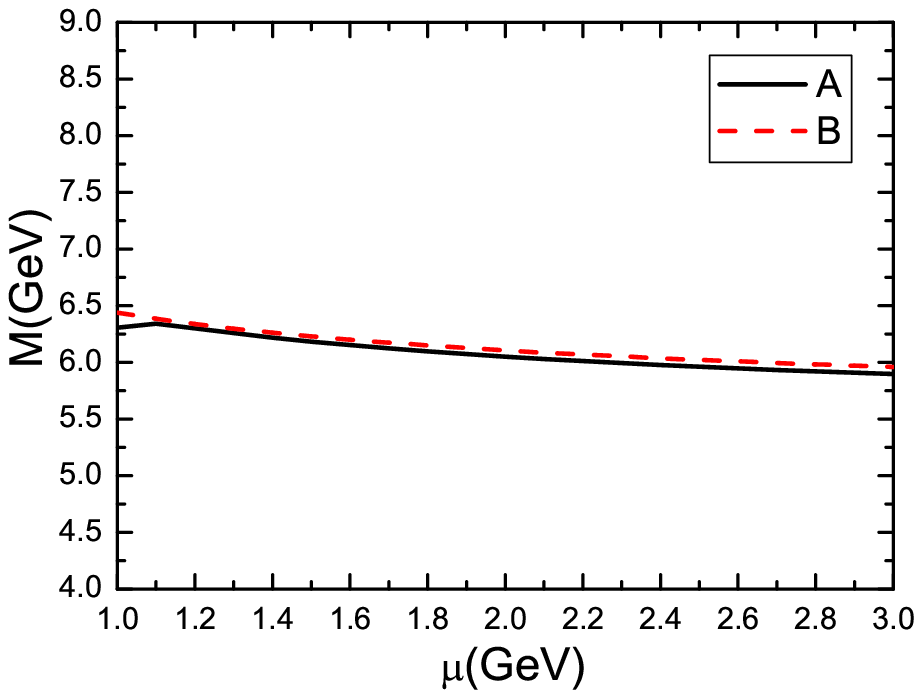}
  \includegraphics[totalheight=5cm,width=7cm]{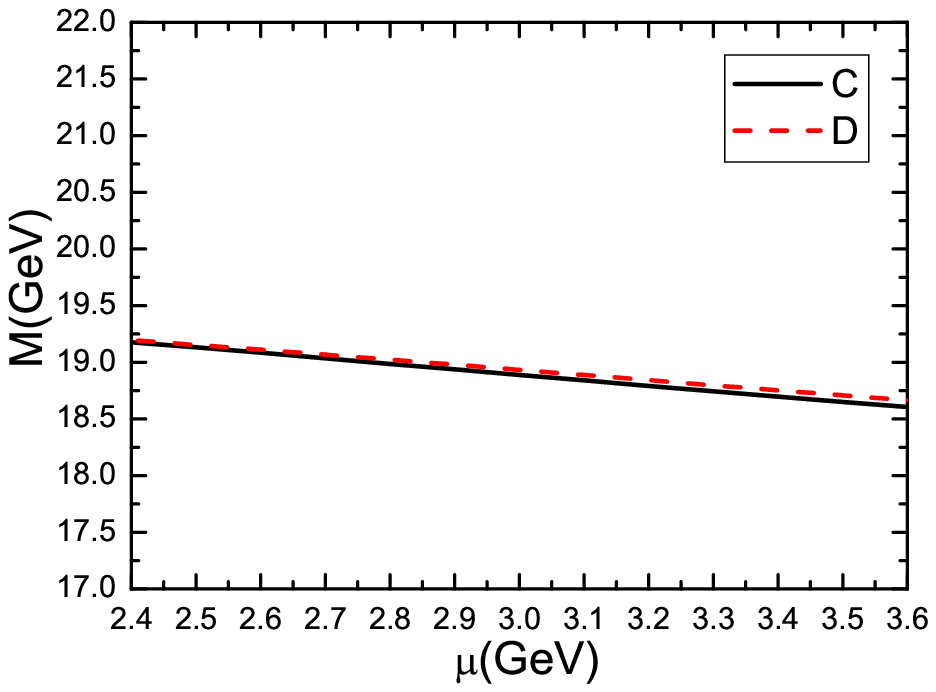}
         \caption{ The masses   of the tetraquark states  with variations of the energy scales $\mu$,  where the $A$, $B$, $C$ and $D$ denote the $cc\bar{c}\bar{c}(1^{+-})$, $cc\bar{c}\bar{c}(1^{--})$, $bb\bar{b}\bar{b}(1^{+-})$ and $bb\bar{b}\bar{b}(1^{--})$, respectively.  }
\end{figure}

\section{Conclusion}
In this article, we construct  the axialvector-diquark-axialvector-antidiquark type  currents to study both the vector and axialvector  $QQ\bar{Q}\bar{Q}$ tetraquark states with the QCD sum rules, and obtain the predictions
$M_{Y(cc\bar{c}\bar{c},1^{+-})} =6.05\pm0.08\,\rm{GeV}$, $M_{Y(cc\bar{c}\bar{c},1^{--})} =6.11\pm0.08\,\rm{GeV}$,
$M_{Y(bb\bar{b}\bar{b},1^{+-})} =18.84\pm0.09\,\rm{GeV}$, $M_{Y(bb\bar{b}\bar{b},1^{--})}  =18.89\pm0.09\,\rm{GeV}$.
The vector tetraquark states lie  $40\,\rm{MeV}$ above the corresponding centroids of the $0^{++}$, $1^{+-}$ and $2^{++}$ tetraquark states, which is a typical  feature  of the  vector tetraquark states consist of four heavy quarks.
 We can search for the $J^{PC}=1^{+-}$ and $1^{--}$ $QQ\bar{Q}\bar{Q}$ tetraquark states in the mass spectrum of the $\mu^+\mu^- +{\rm light\,hadrons}$ and $\mu^+\mu^- \mu^+\mu^-$ respectively in the future.

\section*{Acknowledgements}
This  work is supported by National Natural Science Foundation, Grant Number 11775079.

\end{document}